# TraX Engine: Advanced Processing of Radiation Data Acquired by Timepix Detectors in Space, Medical, Educational and Imaging Applications

**C. Oancea,**[a,1,] **L. Marek,**[a,] **M. Vuolo,**[b] **J. Jakubek,**[a] **E. Soharová,**[a] **J. Ingerle,**[a] **D. Turecek,**[a] **M. Andrlik,**[c] **V. Vondracek,**[c] **, T. Baca,**[c] **M. Sabia,**[d] **R. Kaderabek,**[e] **J. Gajewski,**[f] **A. Rucinski,**[f] **S. Stasica,**[f] **C. Granja,**[a]

[a] *ADVACAM*
  U Pergamenky 12, Prague 7, Czech Republic

[b] *European Space Agency,*
  ESTEC, Keplerlaan 1, NL-2200 AG Noordwijk, Netherlands

[c] *Charles University, Prague, Czech Republic,*
  Prague, Czech Republic

[d] *Eutelsat, OneWeb,*
  London, United Kingdom

[e] *Radalytica,*
  U Pergamenky 12, Prague 7, Czech Republic

[f] *Institute of Nuclear Physics Polish Academy of Sciences*
  PL-31342 Krakow, Poland

  [1]*E-mail*: Cristina.oancea@advacam.cz

[1] Corresponding Author

ABSTRACT:

The TraX Engine is an advanced data processing tool developed by ADVACAM in collaboration with the European Space Agency (ESA), specifically designed for analyzing data from Timepix detectors. This software supports the processing of data from Timepix1, Timepix2, and Timepix3 detectors, which are equipped with various sensor materials (Si, CdTe, GaAs, SiC) and operate in multiple modes (frame-based and data-driven). TraX Engine is capable of processing large datasets across various scientific and medical applications, including space radiation monitoring, particle therapy, and imaging. In space applications, the TraX Engine has been used to process data from satellites like OneWeb JoeySat deployed in LEO orbit, where it continuously monitors space radiation environments measuring flux, dose, and dose rate in real time. In medical applications, particularly in particle therapy, the TraX Engine is used to process data to characterize radiation fields in terms of particle flux, Linear Energy Transfer (LET), and spatial distribution of the radiation dose. The TraX Engine can identify and classify scattered particles, such as secondary protons and electrons, and estimate their contribution to out-of-field doses, a crucial factor in improving treatment planning and reducing the risk of secondary cancers. In imaging applications, the TraX Engine is integrated into Compton cameras, where it supports photon source localization through directional reconstruction of photons. The system's ability to identify the source of gamma radiation with high precision makes it suitable for medical imaging tasks, such as tracking I-131 used in thyroid cancer treatment or localizing radiation sources. This paper presents the architecture and capabilities of the newly developed software TraX Engine, alongside results from various applications, demonstrating its role in particle tracking, radiation monitoring, imaging and others. With its modular architecture, the TraX Engine offers multiple interfaces, including a command-line tool, an API, a web portal and a graphical user interface, ensuring usability across different fields and user expertise levels.

**Contents**

## 1. Introduction

The TraX Engine is an advanced data processing tool developed by ADVACAM s.r.o. within the scope of the Data Processing Engine (DPE) project [1], funded by the European Space Agency (ESA). The TraX Engine, a further development of the DPE software [1], is designed to operate across multiple platforms, including a Command Line Interface (CL), Web Portal (WP), Graphical User Interface (GUI), and Application Programming Interface (API) in C++ and Python. Its primary function is to facilitate both pre-processing and post-processing of data collected by hybrid semiconductor pixel detectors from the Timepix chip family, comprising Timepix1 (TPX) [16], Timepix2 (TPX2) [2], and Timepix3 (TPX3) [3]. The detectors are equipped with various sensor materials, including silicon (Si), cadmium telluride (CdTe), gallium arsenide (GaAs), and silicon carbide (SiC), with thicknesses ranging from 65 µm (SiC) to 2000 µm (CdTe), enabling the detection of a wide spectrum of charged particles. Additionally, with neutron converters, the detectors can detect neutron interactions, with limited efficiency [14-15]. The TraX Engine is compatible with multiple data acquisition modes, including frame-based and data-driven modes, and supports diverse input formats. It produces a range of physics outputs, such as histograms (e.g., deposited energy, cluster size, linear energy transfer (LET), and cluster height), particle flux, dose, and dose rate. Additionally, the engine performs coincidence analysis to identify temporally correlated particle interactions. Advanced post-processing functionalities include directional analysis, which estimates particle track azimuth and elevation angles, and radiation field recognition, which employs significant vector analysis and machine learning algorithms to classify radiation field types by comparison with reference datasets. The TraX Engine incorporates advanced artificial intelligence (AI) machine learning algorithms and heuristic decision trees for particle identification. These algorithms are capable of classifying detected particles into categories such as electrons, protons, ions, and photons. AI models, including neural networks (NN) trained using labelled datasets to perform particle classification offering higher accuracy compared to traditional heuristic methods. This feature is particularly relevant in space radiation environments and medical applications, where precise particle identification is critical for radiation protection and therapy planning. The TraX Engine also enables Compton camera where it supports photon source localization through directional reconstruction of photons. The Timepix detectors have broad applicability across various fields, including imaging with X-rays, protons [4-7], radiation treatment and beam monitoring [8, 23], particle tracking [10-11], particle radiotherapy [9, 12], space radiation monitoring [13], and neutron detection [14-15]. TraX Engine can be used for pre-processing and post-processing data across wide-range of applications where Timepix detectors are used. The aim of this work is to present the TraX Engine as a comprehensive software for advanced data processing in a range of applications, including space radiation monitoring with the OneWeb JoeySat, proton therapy in medical applications, Compton Camera, imaging, accelerator, fundamental and educational physics. The article demonstrates its capabilities in radiation detection, particle identification, and complex data analysis.

## 2. Methods Incorporated in TraX Engine

The TraX Engine provide an extensive set of pre-processing and post-processing algorithms to analyze data from Timepix hybrid semiconductor pixel detectors [1].

### 2.1 Pre-Processing, Detailed Event List

The pre-processing stage of the TraX Engine includes several calibrations and corrections that are essential for ensuring the accuracy of the measured data. The first step is energy calibration, which converts the Time-over-Threshold (ToT) values recorded by the detector into deposited energy. The standard calibration procedure for Timepix detectors is described in [18], with additional details on high-energy calibration provided in [17]. The same principles are applied to the Timepix3 detector [18], while further calibration details for the Timepix2 detector are available in [19-20]. Time-walk correction addresses the variation in time-of-arrival (ToA) caused by differences in signal amplitude. The method implemented for time-walk correction can be found in [22]. Clusterization involves grouping adjacent pixels activated by the same particle event into clusters. These clusters are subsequently analyzed to extract key parameters, such as energy, time-of-arrival, spatial size, and other relevant parameters [12]. In frame-based measurements, clusters are primarily formed based on spatial proximity of the activated pixels, whereas in data-driven mode, both spatial and temporal criteria are used to form clusters [1].

**2.2 Physics products**

**The absorbed dose** in a sensor (e.g., Si, CdTe, SiC, GaAs) quantifies the energy deposited by radiation per unit mass of the sensor material. The dose rate (DR) in the sensor material indicates the rate at which the radiation dose is received over time.

**The particle flux** is calculated as the number of particles passing through a unit area per unit time. In practice, the flux calculation for a radiation detector can be influenced by the geometry of the detector, the efficiency of the detection system, and the specific conditions under which the measurements are taken. Ensuring accurate calibration and correction factors is essential for precise flux determination especially in neutron fields [14-15].

**The Linear Energy Transfer (LET)** quantifies the energy transferred by ionizing radiation to the material through which it passes, per unit length of the material.

The formula for LET can be expressed as:

$LET = \frac{dE}{dx}$,

where **dE** is the cluster deposited energy, energy loss, **dx** is track projected length in $L_{3D}$, **$L_{3D}$** is the length in 3D ($L_{3D}$). It is determined by applying the Pythagoras theorem:

$L_{3D} = \sqrt{(Length\ 2D)^2 + (sensor\ thickness)^2}$, [12, 24]

where $Length_{2D} = L_{2D} - 2.5*\sigma_a$ [24]

The $L_{2D}$ is the projected length in 2D and $\sigma_a$ is the standard deviation along the planar length. The sensor thickness is well known from the production of the sensor and datasheet of the detector. For example, a detector operated at a bias voltage of +200 V fully depletes the sensor volume (Si thickness: $500 \pm 10$ μm).

**Coincidence analysis** is performed by identifying events where two or more particles are detected within a predefined time window. The algorithm tracks temporal correlations between detected particles, allowing for the identification of simultaneous or closely related events (e.g. 100 ns time window). This method is particularly important in multi-detector configurations or experiments with high-energy particle beams, where understanding correlated interactions is essential [1].

**Directional analysis** reconstructs particle trajectories by calculating azimuth and elevation angles based on cluster morphology and time-of-arrival data. This method utilizes spatial and temporal information to accurately determine the particle's path, which is particularly important in applications such as proton therapy and space radiation monitoring, where the precise direction of the particle is a key parameter for treatment planning and environmental assessment [1].

The **spatial mapping** method generates two- and three-dimensional visualizations of key cluster parameters, such as energy deposition and particle flux. These spatial maps are constructed by analyzing the spatial distribution of clusters across the detector's active area. The resulting maps provide detailed radiation field characteristics, enabling applications such as beam profiling, radiation field visualization, and material characterization [1].

**2.3 Algorithms for particle identification**

The particle identification (PID) algorithms implemented in the TraX Engine aim to classify particle clusters by estimating particle type based on their energy and track morphology [1]. Fully connected NN have been developed for the TPX3 detectors with a 500 µm Si sensor and Timepix detector with 300 µm Si sensor, providing robust classification into distinct particle categories. In one configuration, the NN classifies clusters into three primary categories: protons, photons and electrons, and heavier ions [1]. A more advanced configuration utilizes six particle classes, further refining the proton category into low-energy (<30 MeV), medium-energy (30–100 MeV), and high-energy (>100 MeV) protons, while also distinguishing helium ions from other ions. This hierarchical classification enhances the precision of particle identification, making it suitable for complex radiation fields in both space and medical applications. More machine learning algorithms were developed and will be further incorporated in the TraX Engine [21].

**3. Results from Various Applications**

**3.1 Space Radiation: Spacecraft Radiation Monitoring**

For the design and operation of spacecraft, accurately measuring the complex radiation environment in space is essential to assess its impact on sensitive instruments and astronauts. The MiniPIX Space radiation monitor, equipped with a Timepix3 chip, 500 µm thick Si sensor, has been deployed aboard OneWeb's JoeySat satellite in low Earth orbit (LEO), 600 km, since May 2023. This compact device ($90 \times 32 \times 11$ mm$^3$, weighing 140 g) is energy-efficient, consuming less than 3 W. MiniPIX Space continuously monitors space radiation, performing measurements at one-minute intervals. The system is capable of detecting solar events and issuing timely warnings. The adaptive measurement routine implemented on the satellite's computer allows for distinction and analysis of individual particles, even in high-radiation environments. The data collected is processed using the TraX Engine software, which provides a detailed analysis of the radiation field, including its composition, particle directionality, and the quantification of particle radiation effects. As can be seen in Figure 1, radiation field on OneWeb's JoeySat satellite can be decomposed into protons (Fig 1a), electrons and photons (Fig 1b) and heavier ions (Fig 1c). The TraX Engine software enables the extraction of various data products such as particle count rate (Fig. 2 top), flux and dose rate (Fig. 2 bottom), ensuring operational safety for spacecraft systems.

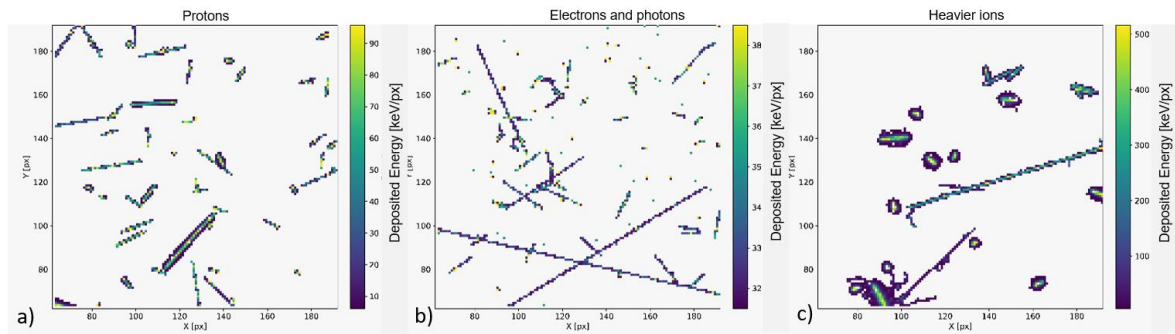

**Figure 1.** Detection and track visualization of space radiation in LEO orbit on the OneWeb JoeySat satellite by the MiniPIX-TPX3 Space payload from Advacam. Radiation components are resolved by Timepix3 with 500 μm Si sensor: a) protons, b) electrons and low-energy gamma rays, c) heavier ions. The energy deposited per pixel is displayed by the color scale. Particle-type classification performed using AI NN.

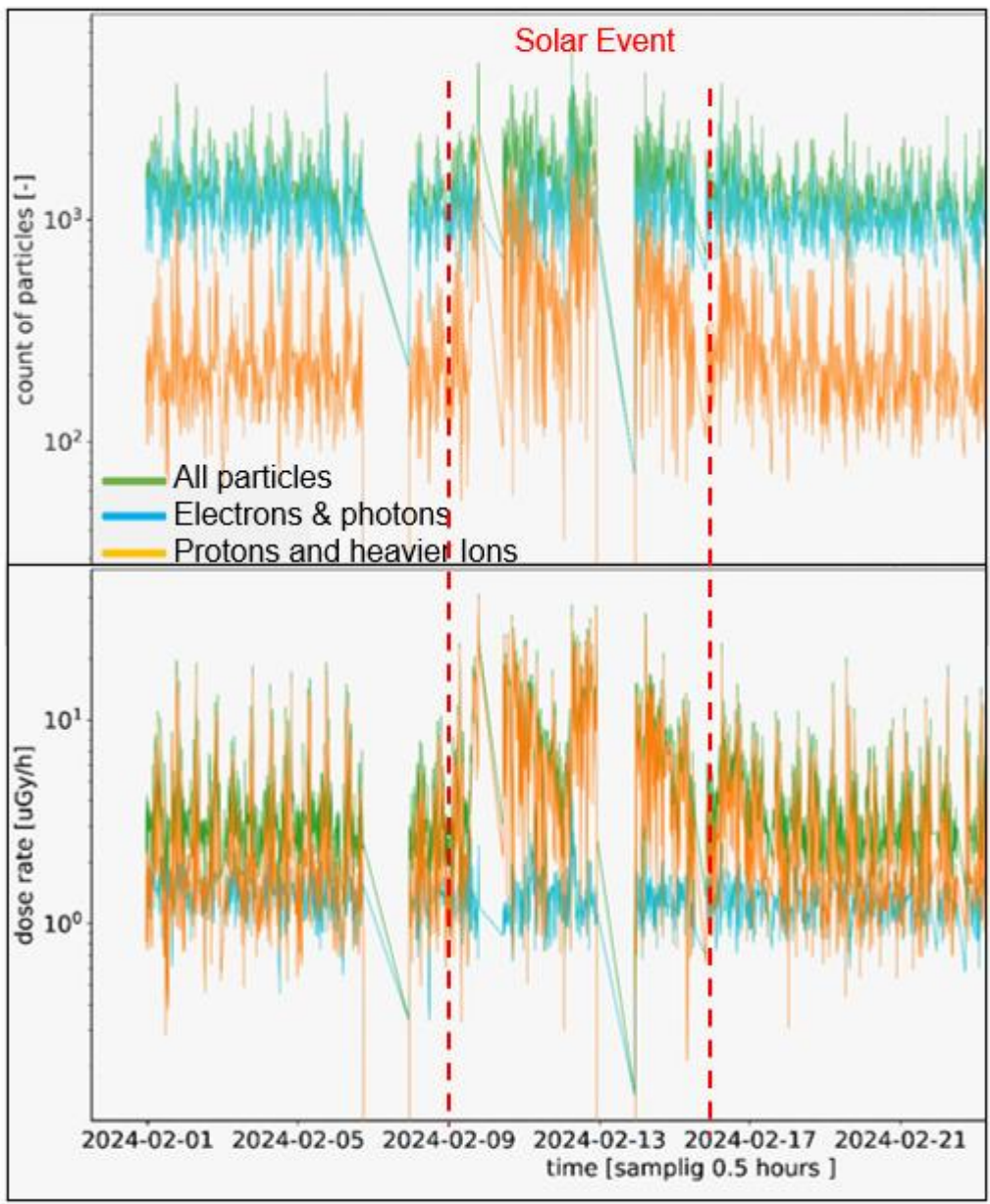

**Figure 2.** Time plot of measured radiation on JoeySat satellite by Timepix3: Particle count (top) and dose rate (bottom) during 25 days (February 2024) including a solar event. Data shown for all particles (green) and broad components: electrons and photons (blue), and protons and heavier ions (orange). Similarly, the dose rate plot shows corresponding variations, with the dose rate increasing significantly during the peak of the event. The dose contributions from protons and heavier ions are more pronounced compared to electrons and photons.

**3.2 Medical: Proton therapy**

Particle therapy is a highly precise external radiation therapy used to treat cancer by delivering particles directly to the tumor site. To ensure optimal treatment outcomes, accurate dosimetry and beam characterization are necessary. In this study, we present the results of experiments using Timepix3 detectors to characterize proton beams, including the measurement of particle flux, energy deposition, and spatial distribution of radiation fields. The experiments were conducted at the Proton Therapy Center Czech (PTC) using a 200 MeV clinical proton beam from a Proteus 235 IBA accelerator. A Timepix3 MiniPIX detector was immersed inside a water phantom to measure the particle field. The water phantom acted as a tissue-equivalent medium, and the detector was positioned at a depth of 15 cm, with 15 cm lateral distances from the Bragg curve. The detector, equipped with a 500 µm silicon sensor, was operated in data-driven mode, collecting both Time-of-Arrival (ToA) and Time-over-Threshold (ToT) data. The main goal of this experiment is to provide characterization of mixed radiation field created around the Bragg curve. This is achieved by the analysis of particle field composition, dose rate, flux, and LET with particle field decomposition performed using TraX Engine software. The particle flux and DR of scattered and secondary radiation can be seen in Figure 3. Although the contribution to the flux is mostly from the class electrons and photons, followed by protons, in the DR can be seen the radiation impact, where protons contribute the most to the dose out-of-field (Fig. 3b).

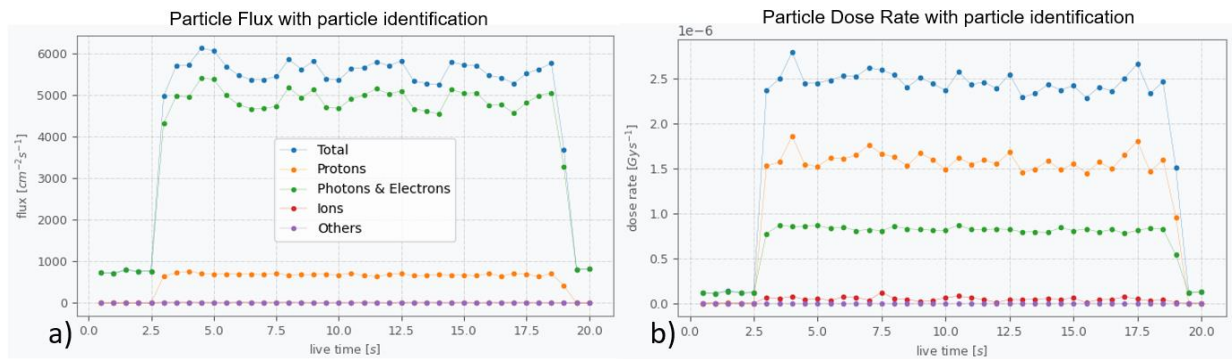

**Figure 3.** a) Particle flux and b) dose rate as a function of live time during proton therapy with detector placed out-of-field. The plots show contributions from various particle types identified and classified using AI NN, including protons, photons and electrons, ions, and other species. Data collection started before beam was turned ON and ended after beam was turned OFF.

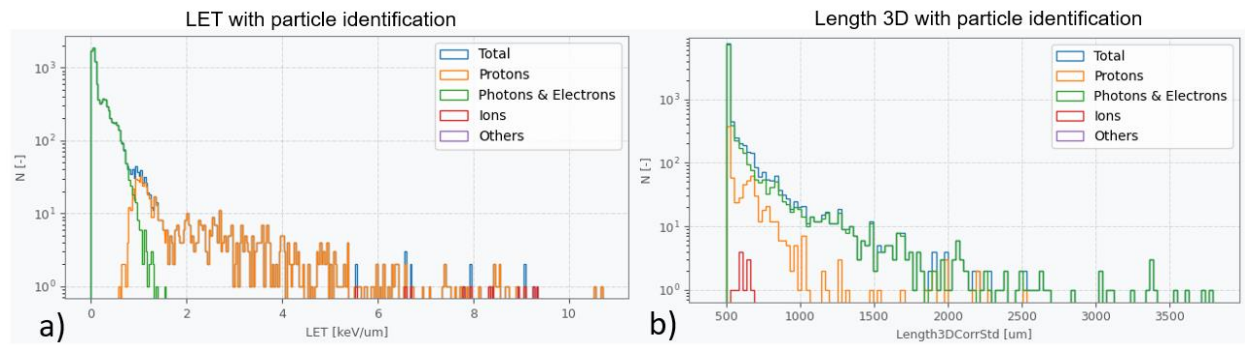

**Figure 4.** a) LET spectra for various particle types measured in proton therapy. The plot shows the distribution of LET values for protons (orange), electrons (green), heavier ions (red), and other species (purple). The total LET distribution (blue) reflects the combination of all particle types. b) Histogram of length in 3D (particle's trajectory through the sensor) used for LET calculation.

LET is a critical parameter for understanding the biological impact of radiation. To derive LET, both the deposited energy in the sensor volume and the length of the particle's path through the sensor (as shown in the histogram of 3D length in Fig. 4b) are used. By using AI NN, we can distinguish the LET spectra for different particles, including protons, electrons, photons, and other ions, as seen in Figure 4a. The low-LET interval is characterized by electrons, with LET values of up to 2 keV/µm, while the high-LET interval is dominated by scattered and secondary protons. Treatment planning systems often focus solely on the primary proton beam, overlooking the contribution of scattered particles to the dose outside the target area. Accurate dosimetry and LET verification are necessary for ensuring precise targeting and minimizing damage to surrounding healthy tissues.

### 3.3 Imaging: Compton camera

The Compton camera developed within the ThyroPIX project is a specialized imaging device designed for precise detection of gamma radiation to monitor thyroid cancer treatments. The system works by detecting scattered photons and reconstructing their origin, allowing for accurate imaging of internal radiation sources. Its integration with advanced pixel detectors like Timepix3 with CdTe sensor material offers significant improvements in imaging performance compared to traditional methods used in medical diagnostics. TraX Engine includes a module that allows the detection of time coincidences in the sensor. Together with information about the deposited energy of particles registered in the detector, it enables the calculation of the original flight direction of the gamma photon, thereby determining the distribution range of the radioisotope.

Figure 5 compares Monte Carlo (MC) simulations and experimental results for I-131 detection using double-layer and single-layer Compton cameras, reconstructed with the TraX Engine's back-projection method. The phantom, shaped like a cross and filled with I-131, simulates the emission source. Results from the double-layer camera show better agreement between simulations and experimental data, with higher clarity and photon detection efficiency compared to the single-layer camera, which demonstrates lower resolution in both simulation and experiment.

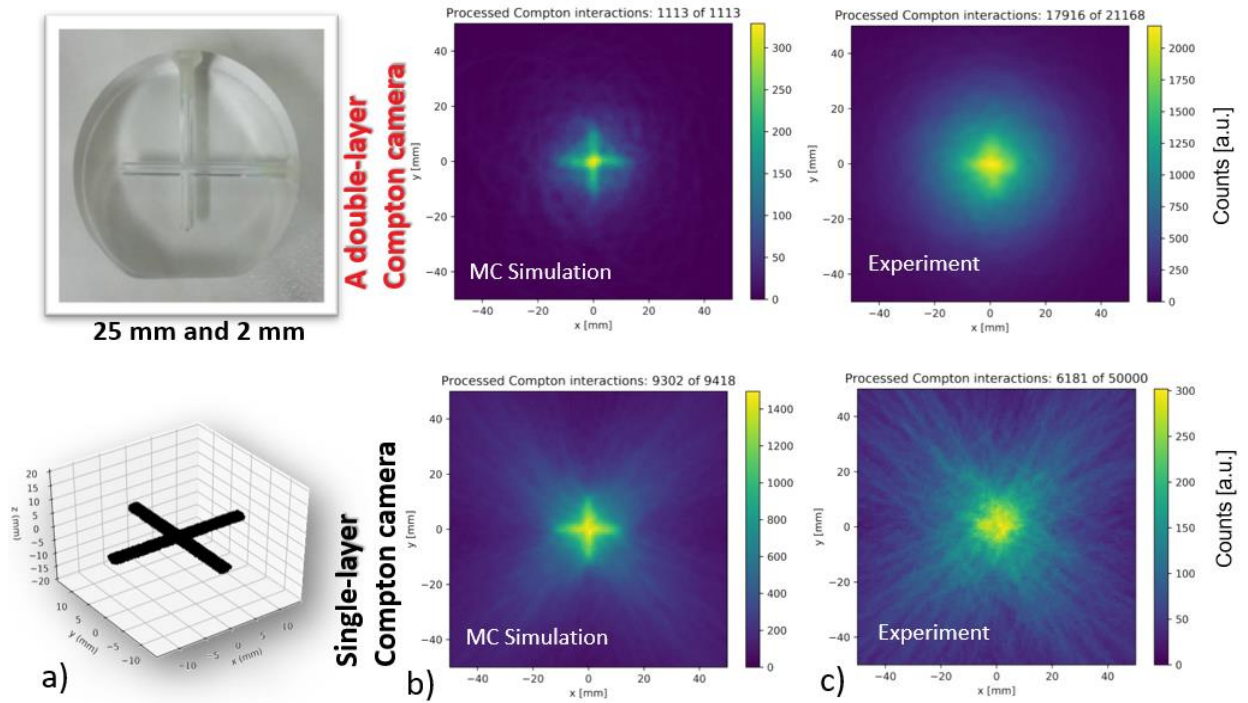

**Figure 5.** Comparison of Monte Carlo (MC) simulations and experimental results obtained using a double-layer and single-layer Compton camera for I-131 detection, reconstructed with the back-projection method incorporated in TraX Engine. a) A phantom shaped like a cross, filled with the I-131 radioisotope (25 mm and 2 mm thick), simulates the emission source for imaging. structure. b) Results from the double-layer Compton camera. c) Results from the single-layer Compton camera, comparing MC simulations (left) with experimental data (right).

### 3.4 Accelerator Physics: beam monitoring, position, direction

In accelerator physics, these detectors are widely used for beam diagnostics, particle tracking, and for assessing beam time structure, dosimetry, temporal characterization, and monitoring [8-12, 23]. Figure 6a shows the beam spot profile of incoming protons at a 45-degree incidence, while Figure 6b illustrates their angular distribution with a measured elevation angle of 45 degrees.

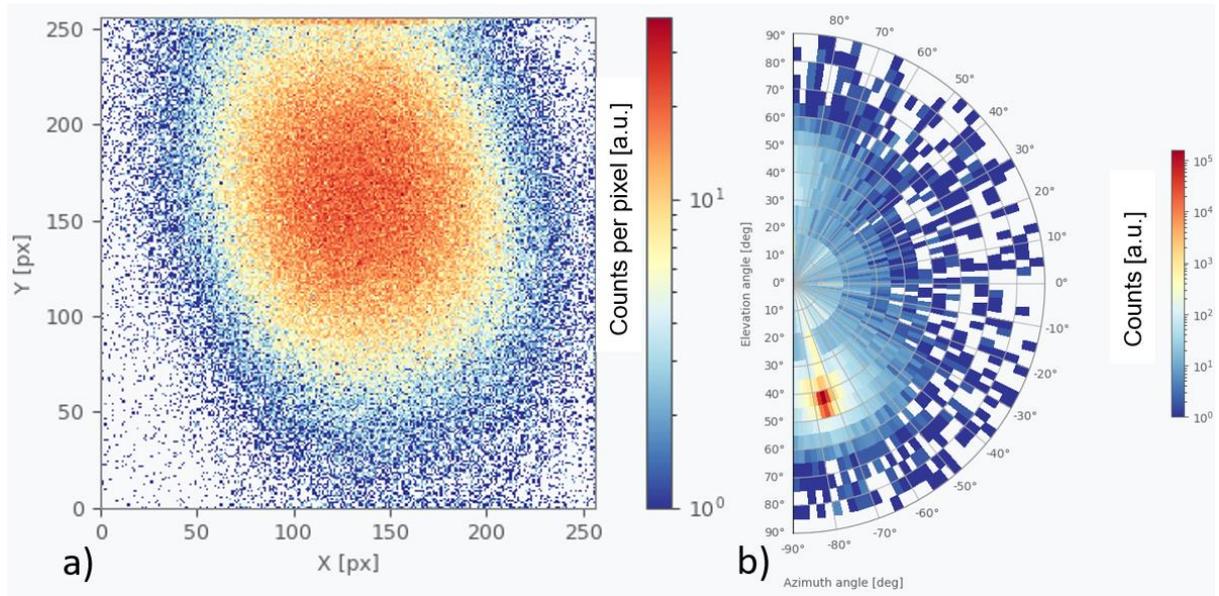

**Figure 6.** a) Map of the proton beam spot profile measured by the Timepix3 detector (Si sensor, 500 μm) in the sensor are of $256 \times 256$ pixels (~2 cm$^2$). b) Directional plot of the incoming particles showing the elevation and azimuth angles. The elevation angle was set at 45 degrees at the detector plane. Data collected at a proton Cyclotron at UJF-PAN, Krakow with 225 MeV proton beams for 5 minutes measurements at low flux. The color scales in both plots represent all particle a) counts per pixel and b) counts per spatial/directional bin displayed in color logarithmic scale.

### 3.5 More applications

In addition to X-ray and gamma-ray imaging, Timepix detectors are also used for charged particle imaging [4-7], providing the potential for sub-pixel spatial resolution. In educational and fundamental physics applications, the TraX Engine simplifies data processing, particularly with its user-friendly graphical interface.

### 4. Conclusions

The TraX Engine is a new data processing tool for Timepix detectors used in various applications. In space, the engine processes radiation data for monitoring spacecraft environments, as demonstrated by the MiniPIX Space monitor on the OneWeb JoeySat. In proton therapy, it enhances dosimetry and beam characterization by single particle identification of scattered radiation, improving accuracy and outcomes. Additionally, in imaging systems, such as the Compton camera, the TraX Engine has improved photon detection and source localization, advancing capabilities in medical diagnostics. The platform's ability to handle large datasets and provide outputs such as particle flux, dose rate, LET, and directional analysis proves essential for understanding and mitigating the effects of radiation in both space and clinical settings. Integration of AI NN for particle identification further enhances its ability to distinguish between different particle species and energy ranges. In conclusion, the TraX Engine sets a new standard for data processing in Timepix detectors, offering comprehensive capabilities for both experts and non-experts across space, medical, imaging, accelerator and fundamental physics applications.

**Supplementary Material**

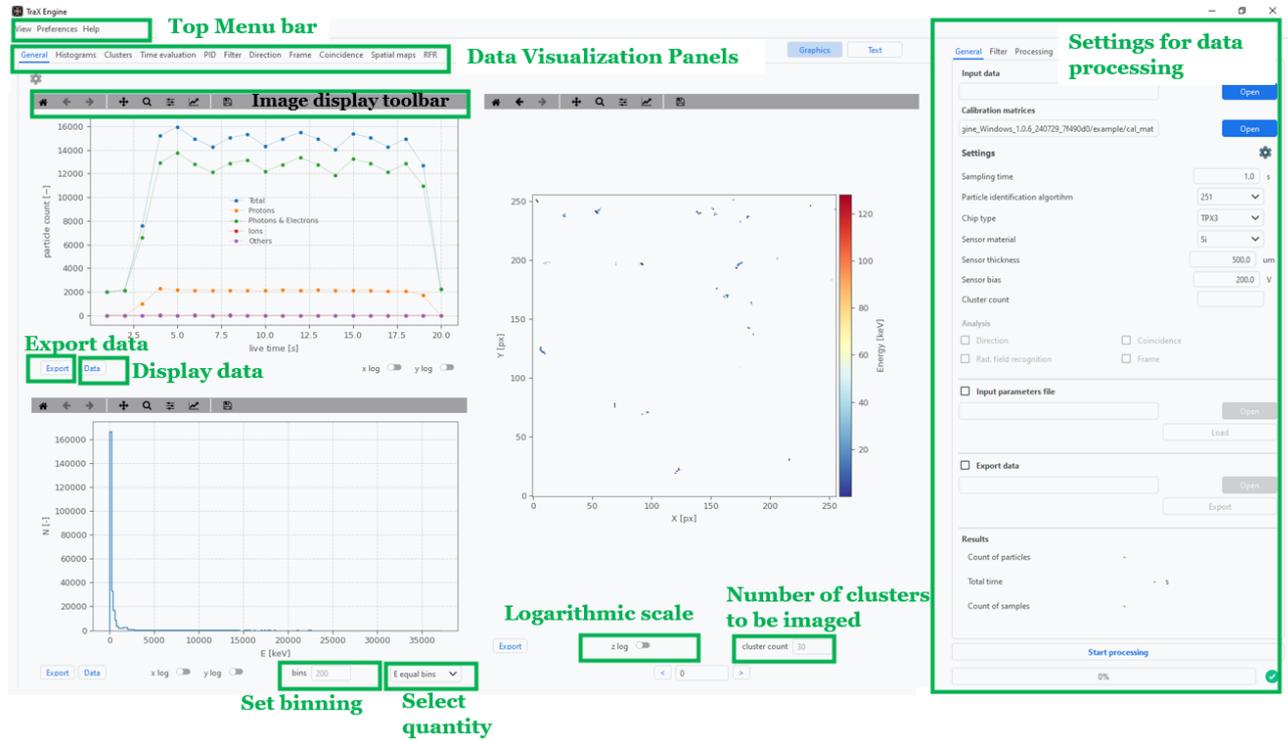

**Figure 7.** Overview of TraX Engine GUI interface - General Panel.